%% file: lvjj_maintext.tex
\RequirePackage{lineno}
\documentclass[aps,prl,twocolumn,showpacs,superscriptaddress,groupedaddress,nofootinbib]{revtex4}

\usepackage{graphicx} 
\usepackage{dcolumn} 
\usepackage{bm}     
\usepackage{amssymb}
\usepackage{multirow}
\usepackage{epsfig}
\usepackage{epstopdf}

\usepackage{amsmath,booktabs,array}

\usepackage{atlasphysics}
\usepackage{eqparbox}
\usepackage{subfigure}

\usepackage{hyperref}

\newcommand{\metpaul}{\mbox{$\rlap{\kern0.15em/}E_T$}}

\newcommand{\WW}{$WW$}
\newcommand{\ZZ}{$ZZ$}
\newcommand{\WZ}{$WZ$}
\newcommand{\alpgen}{{\sc alpgen}}

\newcommand{\jimmy}{{\sc jimmy}}
\newcommand{\mcfm}{{\sc mcfm}}

\newcommand{\photos}{{\sc photos}}
\newcommand{\tauola}{{\sc tauola}}
\newcommand{\mcatnlo}{{\sc mc@nlo}}
\newcommand{\powheg}{{\sc powheg}}
\newcommand{\acermc}{{\sc AcerMC}}
\newcommand{\pythia}{{\sc pythia}}
\newcommand{\calchep}{{\sc calchep}}
\newcommand{\zwprod}{{\sc zwprod}}
\newcommand{\fewz}{{\sc fewz}}

\newcommand{\herwig}{{\sc herwig}}
\newcommand{\sherpa}{{\sc sherpa}}
\newcommand{\geant}{{\sc geant4}}
\newcommand {\Et}   {\ensuremath{E_{\mathrm{T}}}}

\newcommand {\mjj}  {\ensuremath{m_{jj}}}
\newcommand {\mt}   {\ensuremath{m_{\mathrm{T}}}}

\newcommand{\mlvjj}{\ensuremath{m_{\ell\nu jj}}}
\newcommand{\lvlv}{\ensuremath{\ell\nu\ell\nu}}
\newcommand{\evjj}{\ensuremath{e\nu jj}}
\newcommand{\muvjj}{\ensuremath{\mu\nu jj}}
\newcommand{\lvjj}{\ensuremath{\ell\nu jj}}
\newcommand{\lljj}{\ensuremath{\ell\ell jj}}
\newcommand{\llll}{\ensuremath{\ell\ell\ell\ell}}
\newcommand{\lvll}{\ensuremath{\ell\nu\ell\ell}}
\newcommand{\wprime}{\ensuremath{W'}}
\newcommand{\wwz}{\ensuremath{W'~\rightarrow~WZ}}

\newcommand{\graviton}{\ensuremath{G^{*}}}
\newcommand{\gww}{\ensuremath{G^{*}~\rightarrow~WW}}

\newcommand{\siglvjj}{\ensuremath{WW/WZ~\rightarrow~\ell\nu jj}}

\newcommand{\shat}{\ensuremath{\sqrt{s} = 7 \tev}}

\usepackage{preprintcover}
\PreprintCoverPaperTitle{Search for resonant diboson production in the \siglvjj~decay channels with the ATLAS detector at \shat}
\PreprintIdNumber{CERN-PH-EP-2012-296}
\PreprintCoverAbstract{A search for resonant diboson production using a data sample corresponding to $4.7$~\ifb~of integrated luminosity collected by the ATLAS experiment at the Large Hadron Collider in $pp$~collisions at \shat~is presented. The search for a narrow resonance in the \WW~or \WZ~mass distribution is conducted in a final state with an electron or a muon, missing transverse momentum and at least two jets. No significant excess is observed and limits are set using three benchmark models: \WW~resonance masses below 940 \gev~and 710 \gev~are excluded at $95\%$~confidence level for spin-2 Randall--Sundrum and bulk Randall--Sundrum gravitons, respectively; \WZ~resonance masses below 950 \gev~are excluded at $95\%$~confidence level for a spin-1 Extended Gauge Model \wprime~boson.} 
\PreprintJournalName{PRD}

\begin{document}

\title{Search for resonant diboson production in the \siglvjj~decay channels with the ATLAS detector at \shat}

\author{The ATLAS Collaboration}

\date{\today}

\begin{abstract}
A search for resonant diboson production using a data sample corresponding to $4.7$~\ifb~of integrated luminosity collected by the ATLAS experiment at the Large Hadron Collider in $pp$~collisions at \shat~is presented. The search for a narrow resonance in the \WW~or \WZ~mass distribution is conducted in a final state with an electron or a muon, missing transverse momentum and at least two jets. No significant excess is observed and limits are set using three benchmark models: \WW~resonance masses below 940 \gev~and 710 \gev~are excluded at $95\%$~confidence level for spin-2 Randall--Sundrum and bulk Randall--Sundrum gravitons, respectively; \WZ~resonance masses below 950 \gev~are excluded at $95\%$~confidence level for a spin-1 Extended Gauge Model \wprime~boson.

\end{abstract}
\pacs{12.60.Nz, 12.60.Cn}
\maketitle

\section{I. Introduction}

Many extensions to the Standard Model (SM) predict new massive particles that can decay to \WW, \WZ~or \ZZ~final states~\cite{RS1,TC,Higgsless}. In some extensions, the branching ratios of the new particles to these diboson final states greatly exceed their branching ratios to light fermions or photons~\cite{kk_wprime,kkg,kkg_ww}. An analysis of \WW, \WZ~and \ZZ~events is therefore a central element in the search for physics beyond the SM.

This article describes a search for a narrow resonance decaying to either a \WW~or \WZ~diboson intermediate state with subsequent decays to an \lvjj~final state, i.e. a charged lepton (electron or muon), large missing transverse momentum (\met) and at least two jets. Data corresponding to $4.7$~\ifb~collected by the ATLAS experiment at the Large Hadron Collider (LHC) in $pp$~collisions at $\sqrt{s}=7$~\tev\ are used. The search is complementary to other direct searches by the ATLAS Collaboration for a \WW~or \WZ~resonance using events from the \lvlv~\cite{Aad2013860} or \lvll~\cite{ATLAS_Wprime_WZ_lvll}~final state and has the additional advantage of the hadronically decaying $W$~or $Z$~boson in the final state, which leads to a higher branching ratio. Also, the \lvjj~final state allows the reconstruction of the invariant mass of the system, under certain assumptions for neutrino momentum from a $W$ boson decay. Such a reconstruction is not possible in the \lvlv~final state due to the presence of two neutrinos in each event. A separate search for a \ZZ~resonance has been performed using events with a \llll~or \lljj~final state at $\sqrt{s} = 7$~\tev~\cite{ATLAS_ZZ}.

Three benchmark signal models are used to interpret the \lvjj~results. A spin-2 Randall--Sundrum graviton (\graviton) is used to model a narrow resonance decaying to \WW~in two distinct warped extra-dimension models: the original Randall--Sundrum (RS) model~\cite{RS1} (commonly called RS1) and the bulk RS model~\cite{BulkRS_Chang} which allows all SM particles to propagate into the extra dimension. The \WZ~resonance is modeled by a Sequential Standard Model (SSM) \wprime~boson with the $W'WZ$~coupling strength set by the Extended Gauge Model (EGM)~\cite{EGMwprime}. In the EGM model, the $\wprime WZ$~coupling is equal to the SM $WWZ$~coupling strength scaled by a factor $c_{\rm{EGM}} \times (m_{W}/m_{\wprime})^{2}$, producing a partial width proportional to $m_{\wprime}$. In the nominal EGM, the coupling strength scaling factor $c_{\rm{EGM}}$ is set to one. However, this analysis derives exclusion limits for a range of values of this parameter as a function of the invariant mass of the \lvjj~system. This particle is referred to as the EGM \wprime~boson below.

Gauge Model (EGM)~\cite{EGMwprime}. In the EGM model, the $\wprime WZ$~coupling is equal to the SM $WWZ$~coupling strength scaled by a factor $c_{\rm{EGM}} \times (m_{W}/m_{\wprime})^{2}$~with $c_{\rm{EGM}}$~set to 1, thereby producing a partial width proportional to $m_{\wprime}$. This particle is referred to as the EGM \wprime~boson throughout the text.

The aforementioned direct $WW$ resonance search by the ATLAS Collaboration using \lvlv~final-state events in 4.7~\ifb~$pp$~collision data at $\sqrt{s}=7$~\tev~excludes an RS1 graviton with mass less than $1.23$~\tev~and a bulk RS graviton with mass below 840 GeV~\cite{Aad2013860}. Previous searches for a \WW~resonance by the D0 Collaboration in Run II at the Tevatron exclude an RS1 graviton with mass less than 760 GeV~\cite{D0_WW_WZ}. Similar searches, mentioned above, for a \ZZ~resonance by the ATLAS Collaboration exclude an RS1~graviton with mass below 845 GeV~\cite{ATLAS_ZZ}. The CMS Collaboration reports a \ZZ~resonance search in the \lljj~final state and excludes an RS1~graviton with mass below 945 TeV~\cite{cmszz}.  Previous direct searches for a \WZ~resonance at \shat~by the ATLAS and CMS Collaborations exclude the EGM \wprime~benchmark with mass below 760~\gev~\cite{ATLAS_Wprime_WZ_lvll} and 1143~\gev~\cite{CMS_Wprime_WZ_lvll}, respectively. 

\section{II. The ATLAS Detector}

\mbox{ATLAS}~\cite{atlas_detector} is a general-purpose particle detector used to investigate a broad range of different physics processes. Its cylindrical construction is forward-backward symmetric and provides nearly complete hermeticity. The detector is composed of three main subsystems: the inner detector, the calorimeter system and the muon spectrometer. The inner detector (ID) is used for tracking and measuring the momentum of charged particles within the pseudorapidity range $|\eta|<2.5$~\cite{geometry} and is composed of a silicon pixel detector, a silicon microstrip detector and, for $|\eta|<2.0$, a transition radiation tracker. A uniform 2 T magnetic field is provided by a superconducting solenoid surrounding the ID. 
The calorimeter system forms the next layer of the detector, spanning the region $|\eta|<4.9$~and providing three-dimensional reconstruction of particle showers. The inner calorimeter is a high-granularity lead--liquid-argon (LAr) electromagnetic (EM) sampling calorimeter covering $|\eta|<3.2$. Surrounding the EM calorimeter is an iron--scintillator-tile sampling calorimeter providing hadronic coverage in the range $|\eta|<1.7$, extended to $|\eta|<3.2$~with copper--LAr technology. The EM and hadronic calorimeters both have LAr-based forward detectors reaching up to $|\eta|=4.9$. 
Outside the calorimeters, the muon spectrometer (MS) is used to identify muons and measure their momenta. The MS is composed of three large air-core superconducting toroid systems (one barrel and two endcaps) each with eight azimuthally symmetric superconducting coils. Three layers of precision tracking chambers, consisting of drift tubes and cathode strip chambers, allow muon track reconstruction for $|\eta|<2.7$, and fast resistive plate and thin-gap trigger chambers provide trigger signals in the region $|\eta| < 2.4$.

The \mbox{ATLAS} detector uses a three-level trigger system to select events for offline analysis. For this search, events are required to have at least one lepton satisfying trigger requirements, the details of which are presented in section IV.

\section{III. Monte Carlo Simulation}
\label{sec:mc} 

Monte Carlo (MC) simulations are used to model the benchmark signal samples and most SM background
processes.  The RS1 \graviton~and EGM \wprime~boson production and decay are simulated using \pythia~6.4~\cite{pythia} with the modified leading-order (LO$^{*}$) parton distribution function (PDF) set MRST2007LO*~\cite{Sherstnev:2008dm}.
~RS1 \graviton~samples are generated for resonance masses between 500~\gev\ and 1500~\gev\ in 250~\gev\ steps. In these samples the warping parameter, $\tilde{k}\equiv k/\overline{M}_{\rm{Pl}}$,~is set to 0.1, where $\overline{M}_{\rm{Pl}} = M_{\rm{Pl}}/\sqrt{8\pi}$~is the reduced Plank mass. EGM~\wprime~samples are generated with resonance masses from 500~\gev\ to 1500~\gev\ in 100~\gev\ steps, and the production cross-sections are calculated at next-to-next-to-leading order (NNLO) in $\alpha_{S}$~using \zwprod~\cite{zwprod}. The EGM coupling strength scaling factor $c_{\rm{EGM}}$~is set to 1.0 in these samples, which produces a resonance width of $0.032 \times m_{W'}$~\gev.

The bulk RS model is implemented in \calchep~\cite{calchep}, allowing simulation of the $2\rightarrow 4$~production and decay of the graviton with transfer of spin information to the final-state particles. The CTEQ6L LO PDF set~\cite{cteq} is used for these events. Because the bulk RS \graviton~graviton~has 
negligible coupling to light fermions, only gluonic initial states are considered.
These events are processed with \pythia~to simulate the parton shower, hadronization and underlying event. Samples are generated with $\tilde{k}$~of 1.0 and resonance masses from 500~\gev\ to 1500~\gev\ in 100~\gev\ steps, with cross-sections taken from the \calchep\ calculation. For three representative resonance masses, the production cross-sections times branching ratios to $WW/WZ$~for each sample are given in Table~\ref{t-signal-xs}.

\newcolumntype{d}{D{,}{\,.\,}{-1,-1}}

\begin{table}
  \caption[SignalCrossSections]{Production cross-sections times  branching ratios for~\gww~or \wwz~for the RS1 \graviton, bulk RS \graviton, and the EGM \wprime, 
for resonance masses equal to 500~\gev, 1000~\gev, and 1500~\gev. All cross-sections are given in picobarns.}
  \begin{center}{\small
  \begin{tabular}{l | c c c}
    \hline
    \hline
    &{\bf RS1 \graviton} & {\bf Bulk RS \graviton} & {\bf EGM \wprime} \\ \hline
    Mass [\gev]& \multicolumn{3}{c}{$\sigma~\times$ BR [pb]}\\
    \hline

    500     &  13.0    & 3.0    & 2.6 \\
    1000    & 0.23     & 0.023    & 0.10 \\
    1500    & 0.017  & 0.0011    & 0.011 \\

    \hline
    \hline
  \end{tabular}}
    \label{t-signal-xs}
\end{center}
\end{table}

\newcolumntype{d}{D{,}{\,\pm\,}{-1,-1}}

Templates with $50$~\gev\ spacing in the mass of the \lvjj~system, \mlvjj, are constructed to ensure a signal prediction if no signal MC sample is generated at that mass. These templates are created by first fitting the fully simulated \mlvjj~distribution with a Crystal Ball function~\cite{crystal_ball}. The shape parameters from these fits are interpolated across the mass range $500\mbox{--}1500$~\gev~and used to construct Crystal Ball functions, the signal templates, at the intermediate mass points. The acceptances for these signal templates are also interpolated from fits to the acceptances of the fully simulated samples.

For SM background processes, the production of a $W$~or $Z$~boson in association with jets is simulated with \alpgen~\cite{alpgen} using the CTEQ6L LO PDF set.  These events are processed with \herwig~\cite{herwig}~for parton showering and hadronization, and \jimmy~\cite{jimmy}~to 
simulate the underlying event. The samples are initially normalized to the NNLO production cross-section computed 
with \fewz~\cite{fewz,fewz2}.  The prediction of the $W$~boson transverse momentum, \pt, by \alpgen~is reweighted to agree with the shape predicted 
by \sherpa~\cite{sherpa}, which is observed to agree more closely with data at high \pt~\cite{atlas_wjets}.
Single top quark ($tb,~tqb,~tW$) and top quark pair (\ttbar) production are simulated with  
the next-to-leading-order (NLO) generator
\mcatnlo~\cite{mcatnlo,mcatnlo_singletop,mcatnlo_singletop2} interfaced to \herwig~and \jimmy~and using the
CT10~\cite{Lai:2010vv} NLO PDF set. A sample of \ttbar~events generated with 
\powheg~\cite{powheg1,powheg2,powheg3}~interfaced to \herwig~and \jimmy~is used to cross-check 
the \mcatnlo~\ttbar~production model, and a \powheg~\ttbar~sample interfaced to \pythia~is generated to 
study the dependence on the parton shower and hadronization model.
The \acermc~event generator~\cite{acer}~interfaced with \pythia~is employed to study the effect of initial- and final-state radiation in \ttbar~events. Both $\ttbar$~and single top quark samples are generated assuming 
a top quark mass, $m_{t}$, of $172.5$~\gev, but two \mcatnlo~\ttbar~samples are generated with $m_{t}=170$~\gev~and 
175~\gev\ to determine the dependence of the background prediction on the top quark mass.
The \ttbar~cross-section is normalized to the approximate NNLO value~\cite{ttbar_Moch,ttbar_Langenfeld}. Single top quark production cross-sections are taken from an NNLO calculation for the $tb$~process~\cite{tb}, and approximate NNLO calculations for the $tqb$~and $tW$~processes~\cite{tqb_tW}. SM diboson production ($WW, WZ, ZZ$) is modeled using \herwig~and normalized to the NLO production cross-sections computed by \mcfm~\cite{mcfm,dibosons}~with the MRST2007LO*~PDF set. In all samples, \photos~\cite{photos}~is employed to simulate final-state photon radiation and \tauola~\cite{tauola}~to take into account polarization in $\tau$~lepton decays.

All MC samples include the effect of multiple $pp$~interactions (pile-up) per bunch crossing and are reweighted so as to match the distribution of the number of interactions per bunch crossing to that observed in the data. The detector response is simulated using a \geant-based model~\cite{geant} of the ATLAS detector~\cite{atlas-sim}. Finally, events are reconstructed using the same software used for collision data. 

\section{IV. Object Reconstruction and Event Selection}
\label{sec:event_selections}

The events recorded by the ATLAS detector for this analysis are selected by single-electron or single-muon triggers. The electron trigger requires an electron-like object~\cite{ATLAS_triggerops} with transverse energy (\Et) greater than 20~\gev\ or 22~\gev\ depending on the LHC instantaneous luminosity. The muon trigger requires a muon candidate with \pt~$ > 18$~\gev. The data sample used, collected in 2011, corresponds to an integrated luminosity of $4.7$~\ifb~\cite{ATLAS_lumi,ATLAS_lumi2} after applying data-quality requirements~\cite{2012EPJC.72.1960G}. MC events must satisfy the same trigger selection requirements.

All triggered events must have at least one reconstructed vertex formed by the intersection of at least three tracks with \pt~$ > 400$~MeV~\cite{ATLAS_tracking}. From the list of all vertices satisfying this requirement, the vertex with the largest sum of squared \pt~of the associated tracks is assumed to be the primary hard-scatter vertex (PV).

Electrons are reconstructed from energy clusters in the calorimeter with an electromagnetic shower profile
consistent with that expected for an electron, and must have a matching ID track. 
Electron candidates must have $E_{\rm{T}} > 30$~\gev\ and be found within the fiducial region defined by
$|\eta|<2.47$, excluding the region $1.37 < |\eta| < 1.52$~which corresponds to the poorly instrumented
transition between the barrel and endcap calorimeters.
The longitudinal impact parameter of the electron track with respect to the PV ($|z_{0}|$) must be less than $1$~mm, and 
the significance of its transverse impact parameter with respect to the PV
($|d_{0}|/\sigma_{d_{0}}$) must be less than $10$. 

Electron candidates must also be isolated from other activity in the calorimeter, such that the sum of calorimeter transverse energy in a cone of radius~$\Delta R = \sqrt{(\Delta \phi)^2 + (\Delta \eta)^2} = 0.3$ around the electron, corrected for pile-up contributions and the electron energy, is less than $6$~\gev. 
The energy scale and resolution for electrons in MC events are corrected to match that in $Z\rightarrow e^{+}e^{-}$~events~\cite{ATLAS_electron} measured in data.

Muons are reconstructed from the combination of tracks formed from hits in the MS and the ID~\cite{ATLAS_muon,ATLAS_muon2}.
The combined muon track must have \pt~$ > 30$~\gev\ and $|\eta|<2.4$. The muon track must have $|z_{0}| < 10$~mm and $|d_{0}|/\sigma_{d_{0}} < 10$. The difference in $|z_{0}|$~requirements between the electron and muon tracks results from the higher fraction of misreconstructed electrons due to QCD multi-jet events.

Furthermore, muon candidates must be isolated from other tracks and calorimeter activity: the sum of track transverse momenta surrounding the muon track in a cone of radius $\Delta R = 0.3$ must be less than $15\%$~of the muon \pt; the calorimeter transverse energy, corrected for pile-up contributions, in a cone of radius $\Delta R = 0.3$~must be less than $14\%$~of the muon \pt. The muon \pt~scale and resolution in MC events are adjusted to match that in $Z\rightarrow \mu^{+}\mu^{-}$~events measured in data~\cite{ATLAS_muon3}.

Jets are reconstructed using the anti-$k_{t}$~sequential recombination clustering algorithm~\cite{Cacciari:2005hq,cacciari-2008}, with radius set to $0.4$. The inputs to the reconstruction algorithm are topological energy clusters~\cite{jesjer} calibrated at the EM energy scale, appropriate for the energy deposited by electrons or photons~\cite{jesjer}. These jets are then calibrated to the hadronic energy scale, using $\pT$- and $\eta$-dependent correction factors obtained from simulation. The uncertainty on these correction factors is determined from control samples in data. Jets originating from the PV are selected by requiring that at least 75\% of the \pt\ sum of tracks matched to the jet belongs to tracks originating from the PV. If a reconstructed electron and jet candidate overlap within $\Delta R=0.3$, the jet is rejected. Finally, jets must have \pt~$>40$~\gev\ and $|\eta|<2.8$.

Jets originating from $b$-quarks are identified by exploiting the long lifetimes 
of bottom hadrons, which lead to observable decay lengths in the detector.
The $\rm{SV0}$~secondary vertex $b$-tagger~\cite{ATLAS_btag,ATLAS_btag2} is used at an 
operating point yielding an average $b$-jet-tagging efficiency of 50\% in simulated 
\ttbar\ events and an average light-quark jet rejection factor of 200.

The missing transverse momentum (\met) is defined as the negative vector sum of transverse energies 
or momenta of all objects in the event. The ATLAS \met~algorithm~\cite{ATLAS_metalgo}~combines the \pt~of muons reconstructed in the MS with the transverse energies measured in calorimeter cells associated either to physics objects (such as jets or leptons) or to topological clusters not associated with physics objects. Calorimeter cells used in the \met~calculation are calibrated individually according to the physics object to which they are associated. Cells in topological energy clusters that are not associated with any reconstructed high-\pt~object are calibrated separately using the local hadronic calibration scheme~\cite{ATLAS_lcw}.

In the initial selection, events must contain exactly one electron or muon, and must have \met~$> 40$~\gev. Events are also required to contain at least two jets, with the requirement that the highest-\pt~jet has \pt~$>100$~\gev. In the following, events with an electron are labeled \evjj~and muon events are labeled \muvjj. To reduce the QCD multi-jet background, two triangular veto regions are constructed in the plane defined by the \met~and $\Delta\phi(\ell,\met)$, the difference in azimuthal angle between the lepton and \met directions. The first region, defined by \mbox{$|\Delta\phi| < 1.5 - 1.5\times(\met/75~\rm{GeV})$},~corresponds to events where the lepton and \met~directions are aligned. Back-to-back event topologies populate the second region defined by \mbox{$|\Delta\phi| > 2.0 + (\pi-2)\times(\met/75~\rm{GeV})$}. Events falling in either of these two regions are rejected. The selection cuts described above define the preselection criteria.

\section{V. Background Estimation}
\label{sec:background}

Background sources are classified into two categories based on the origin of the charged lepton in the event.
The first category includes backgrounds where the charged lepton is produced in the decay of a $W$~or $Z$~boson.
The second category corresponds to all other sources, including both events with a misidentified lepton, e.g.\ where
a jet with a large electromagnetic energy fraction passes the electron selection requirements, and events 
with a true lepton produced in a hadron decay.

Backgrounds from the first category, which include $W/Z$~+ jets, \ttbar, single top quark, and diboson production, 
are modeled with MC events and are normalized to the product of the production cross-section for that background 
and the total integrated luminosity of the dataset. The normalization of the $W$ + jets and \ttbar~backgrounds is further tested using data as described in Section VI.

Backgrounds in the second category are modeled with independent samples of collision data based on the following prescriptions.  In the \evjj\ channel, the sample is selected by inverting the calorimeter isolation requirement for electron candidates that satisfy all other selection criteria. This selects events that are likely to originate from multi-jet production, but have kinematic properties that are very similar to those multi-jet events that pass the isolation requirement. In the \muvjj\ channel, the primary source of these backgrounds are semileptonic decays of hadrons within a jet. Events with muons that satisfy all selection criteria except the transverse impact parameter significance cut are used to model this background. Kinematic variable templates are derived from these samples after subtracting the contributions from backgrounds in the first category.

The data-driven backgrounds in the second category, henceforth labeled ``fake'' lepton backgrounds, are then normalized together with the $W$~+ jets background through a likelihood fit to the data in a region with negligible signal contamination. This is done separately for the \evjj\ and \muvjj\ channels using the lepton transverse mass distribution, $\mt \equiv \sqrt{2\pt^{\ell}\met(1-\cos(\Delta\phi))}$, which distinguishes events with charged leptons from a $W$~boson decay from events with a ``fake'' lepton. The normalization of all other backgrounds, from the first category, remains fixed in the fit. 

The distributions of the lepton \pt, \met~and the leading jet \pt~in data and for the predicted backgrounds, after applying the event preselection criteria, are shown in Fig.~\ref{f-preselection}. In this figure, the associated errors are combination of the systematic and statistical uncertainties.
Table~\ref{t-preselection-yields}~shows the yields for each background and for the data. The total estimated background and the data agree within the expected total uncertainty at this stage of the selection.

\begin{figure*}
  \centering
  \subfigure[]{\includegraphics[width=0.32\textwidth]{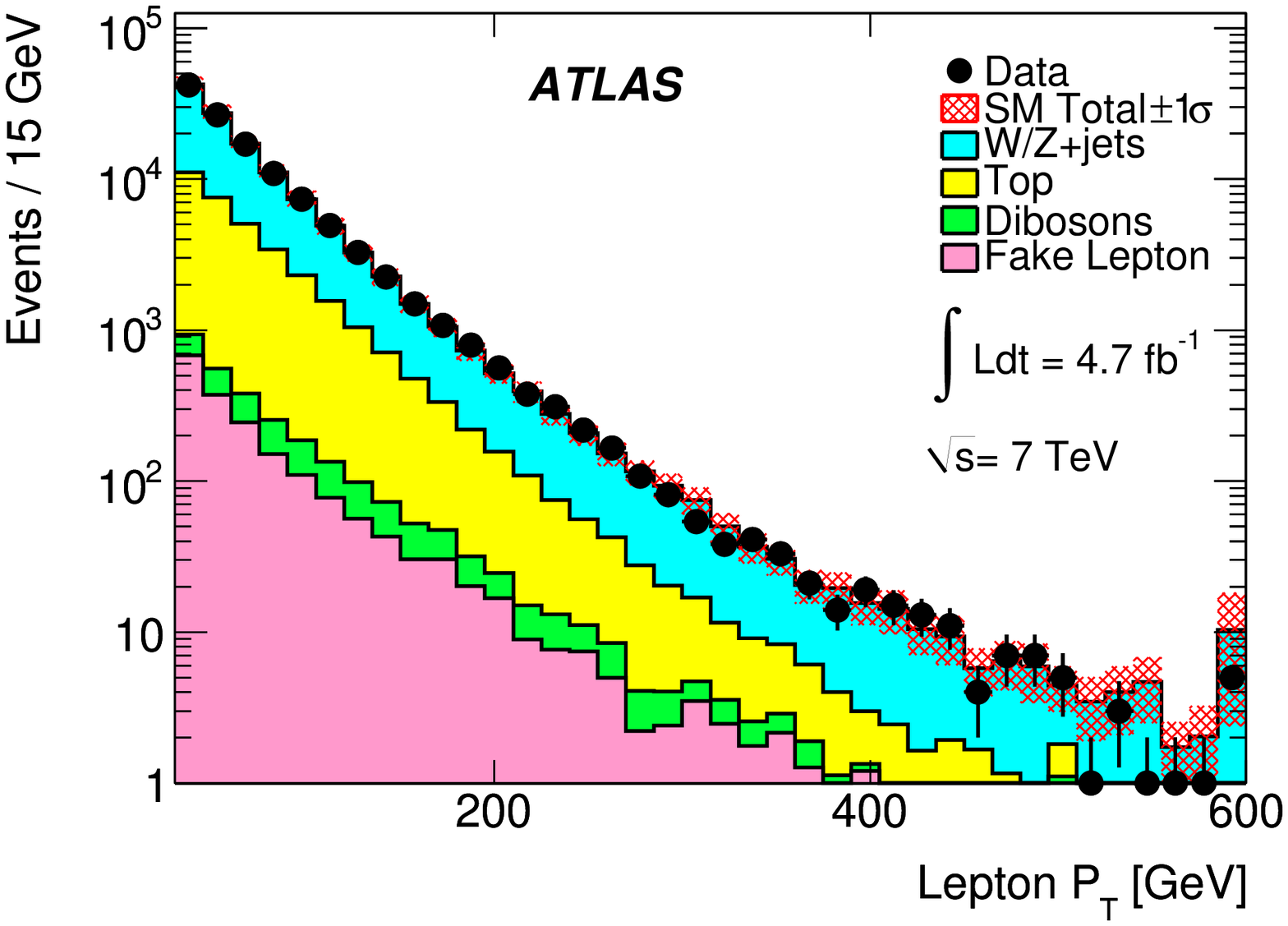}}
  \subfigure[]{\includegraphics[width=0.32\textwidth]{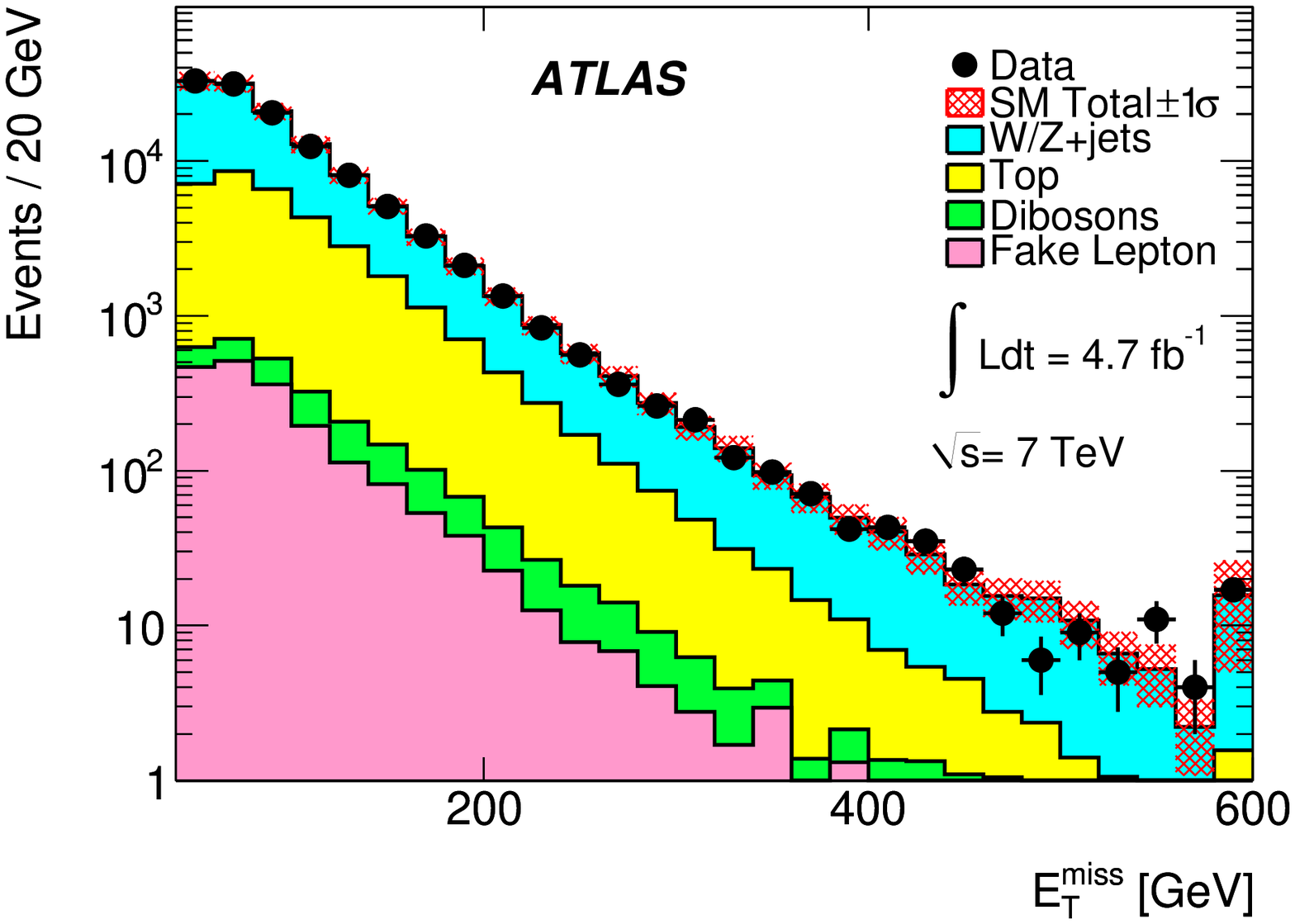}}
  \subfigure[]{\includegraphics[width=0.32\textwidth]{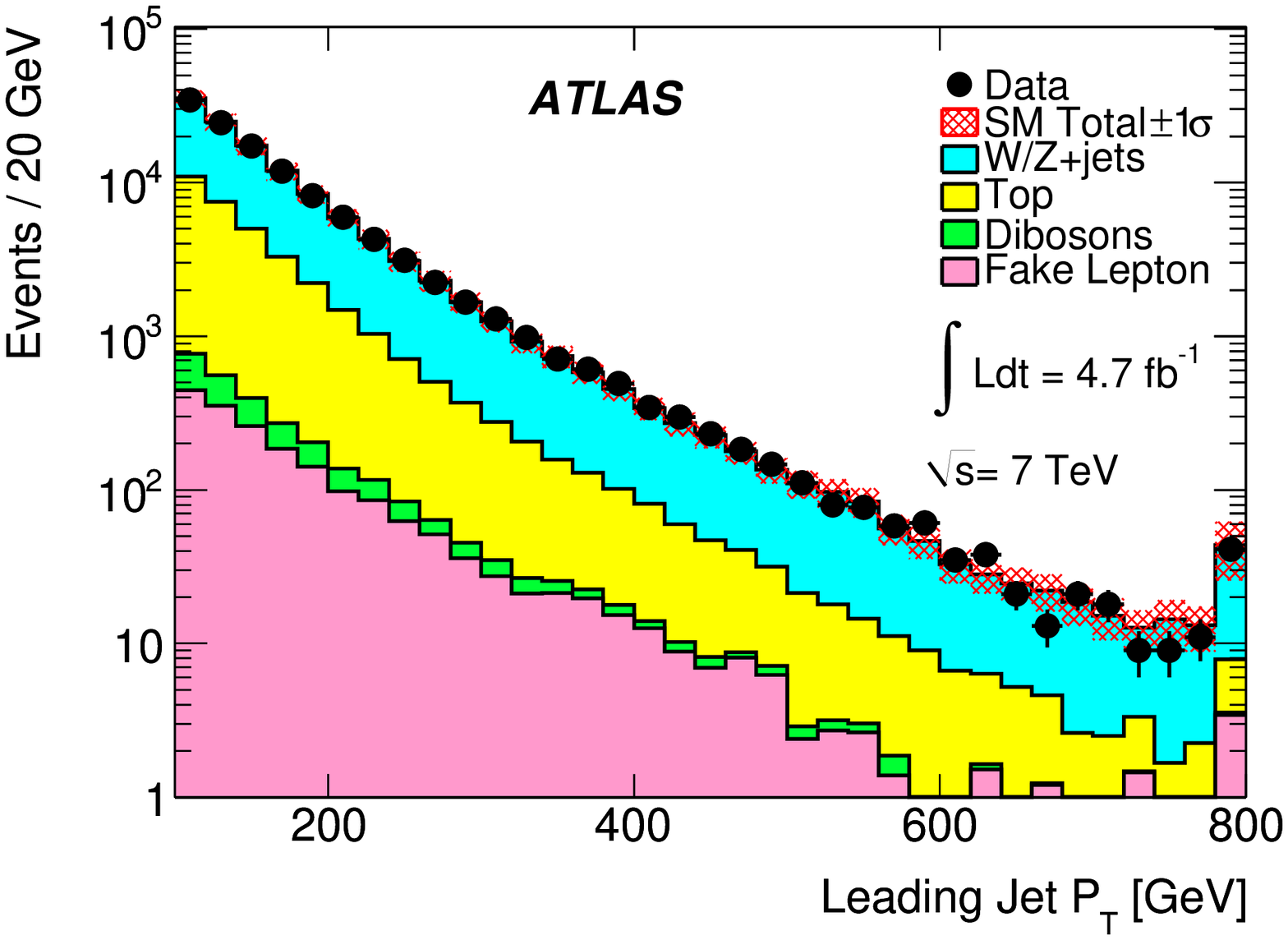}}
  \caption{\label{f-preselection}
    (color online) Data and background predictions for (a) the lepton \pt, (b)~\met, and (c)~leading jet \pt~for preselected events. Electron and muon events are combined in all plots. The right-most bin contains overflow events.}
\end{figure*}

\begin{table}
  \caption{\label{t-preselection-yields} The number of data and estimated background events after applying the preselection cuts. The associated errors are the quadrature sum of the systematic and statistical uncertainties.}
  \begin{center}{\small
    \begin{tabular}{l | d | d}
      \hline
      \hline
      Process &\multicolumn{1}{c|}{\bf \evjj~channel} &\multicolumn{1}{c}{\bf \muvjj~channel} \\
              &\multicolumn{1}{c|}{\bf [$10^3$ events]} &\multicolumn{1}{c}{\bf[$10^3$ events] }\\
      \hline
      $W$+jets & 38.0,4.1 & 46.0, 4.8\\ 
      $Z$+jets & 1.3 , 0.1 & 1.8,0.2\\ 
      Top & 15.0,1.4 & 16.0,1.5\\
      Diboson & 0.5,0.1 & 0.5,0.1\\
      Fake lepton & 0.9,0.2 & 0.5,0.2\\
      Total background & 56.0,4.3 & 65.0,5.1\\
      Data & 55.2  & 64.2 \\
      \hline
      \hline
    \end{tabular}}
  \end{center}
\end{table}

\section{VI. Selection of Signal and Control Regions}
\label{sec:signalselection}

The \WW\ or \WZ\ mass, \mlvjj, is calculated as the invariant mass of the $\ell\nu jj$ system. To reconstruct this quantity, the $x$~and $y$~components of the neutrino momentum vector, \px~and \py, are set equal to $\met\cos(\phi_{\rm{miss}})$~and $\met\sin(\phi_{\rm{miss}})$, 
respectively, with $\phi_{\rm{miss}}$~corresponding to the direction of the \met~vector~in the tranverse plane. The neutrino \pz~is obtained by imposing the $W$~boson mass constraint in the momentum conservation equation. It is defined as either the real component of the complex \pz~solution or the minimum of the two real solutions. 
In events with three or more jets, the two jets with the highest transverse momenta are considered.

In signal events, the \pt~of each boson peaks near half of the resonance mass, and the dijet mass distribution, \mjj, is characterized by a peak close to the $W$~or $Z$~boson mass. 
Since this analysis searches for resonant masses larger than $500$~\gev, the signal region is  defined by requiring the reconstructed \pt~of the dijet system and of the lepton--\met~system to be greater than $200$~\gev~and the reconstructed dijet mass to be within the window $65<\mjj<115$~\gev. Figure~\ref{f-final-mjj} compares the \mjj~distribution observed in data with those predicted for the backgrounds and an enhanced EGM \wprime~signal after the requirements on the reconstructed dijet and lepton--\met~\pt~values but without the dijet mass window requirement.

\begin{figure}
  \centering
  \includegraphics[width=0.45\textwidth]{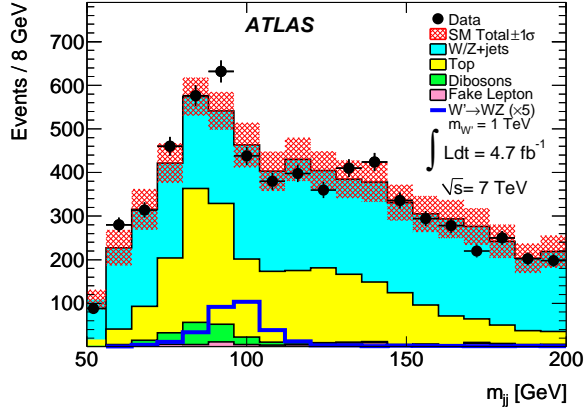}
  \caption{(color online) Observed and predicted \mjj~distribution in all events satisfying the \pt~selection requirements of the reconstructed $W/Z$~bosons. Predictions for an EGM \wprime~boson, with the signal cross-section enhanced by a factor of five, are shown for a resonance mass of 1 \tev.
  \label{f-final-mjj}}
\end{figure}

Two control regions are created to test the $W$ + jets and \ttbar~background modeling of the \mlvjj~distribution.
The $W$ + jets control region is identical to the signal region, except for the \mjj~requirement, which is inverted. Two independent sidebands are formed, $\mjj<65$~\gev~and $\mjj>115$~\gev. A scale factor, defined as the number of data events divided by the total background prediction, is computed in each sideband and parameterized as a function of \mlvjj. 
The weighted average of the scale factors, found in the $\mjj<65$~\gev~and $\mjj>115$~\gev~sidebands, has a value of 1.012 and is used to normalize the $W$~+ jets background prediction in the signal region. The difference between the individual scale factors is used as the uncertainty on this normalization. The two sidebands are combined in Fig.~\ref{f-wjets-cr}, which shows the \mlvjj~distribution for the $W$~+ jets control region after applying the $W$~+ jets scale factors. Good agreement between the data and MC is observed.

\begin{figure}
  \centering
  \includegraphics[width=0.45\textwidth]{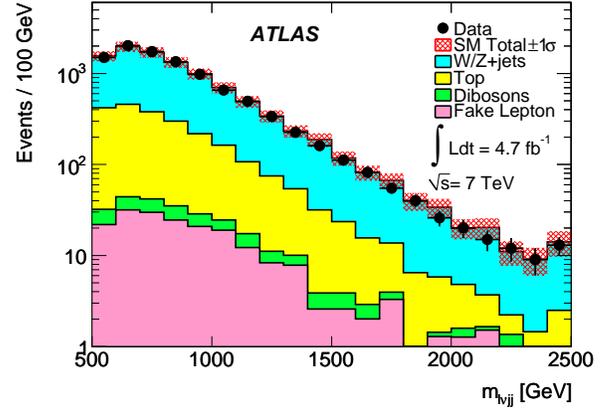}
  \caption{\label{f-wjets-cr}
    (color online) The \mlvjj~distribution for the data and the background predictions for events in the $W$~+ jets~background control region. The right-most bin contains overflow events.}
\end{figure}

The \ttbar~control region is created by selecting events with at least two $b$-tagged jets. The reconstructed \pt~of the dijet system is required to be greater than $200$~\gev, and events are required to have $\mjj<65$~\gev~or $\mjj>115$~\gev~to avoid overlap with the signal region. Figure~\ref{f-ttbar-cr} shows \mlvjj~for all events in the \ttbar~control region. In this control region, $587 \pm 87$~\ttbar~events and $42 \pm 6$~events from other backgrounds are expected and $602$ data events are observed. Given the agreement observed in the \ttbar~control region, no normalization correction is applied to the \ttbar\ background prediction in the signal region.

\begin{figure}
  \centering
  \subfigure{\includegraphics[width=0.40\textwidth]{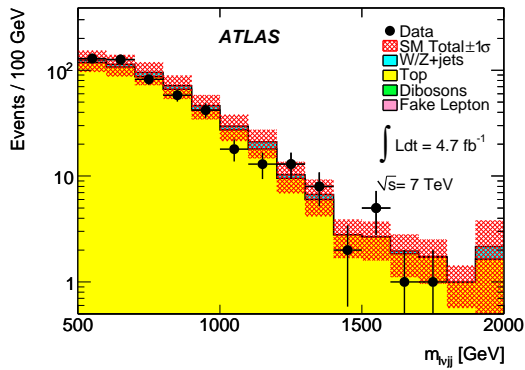}}
\caption{\label{f-ttbar-cr}
    (color online) The \mlvjj~distribution in data events and the estimated backgrounds for the \ttbar~background control region. The right-most bin contains overflow events.}
\end{figure}

\section{VII. Systematic Uncertainties}
\label{sec:systematics}

Systematic uncertainties that affect the predicted signal acceptance and background rate are grouped into three independent categories: uncertainties due to the limited precision of theoretical calculations, experimental uncertainties on the event reconstruction efficiencies and resolutions, and the determination of the integrated luminosity. Uncertainties from the first and third categories impact the signal and all of the backgrounds except $W$ + jets and ``fake'' lepton backgrounds which are estimated from data. The integrated luminosity uncertainty is~$3.9\%$~\cite{ATLAS_lumi,ATLAS_lumi2}.

Several sources of theoretical uncertainty on the \ttbar~background rate are considered. The largest of these is the $^{+7}_{-10}\%$~\cite{ttbar_Moch,ttbar_Langenfeld} uncertainty on the production cross-section. Additionally, the magnitudes of the following systematic uncertainties affecting the \ttbar~background distribution vary with \mlvjj. The largest deviation from the \ttbar~prediction for all \mlvjj~values is presented below. The nominal \mcatnlo~model for \ttbar~production~differs from the \powheg~model by at most $3\%$.  A $1\mbox{--}2\%$~variation is measured when the top quark mass is varied by $\pm2.5$~\gev~using \mcatnlo~MC~samples. The difference between the nominal \herwig~parton shower model and the \pythia~model in \powheg~generated events is at most $2\%$. Finally, the uncertainty due to the initial-state radiation (ISR) and final-state radiation (FSR) model in \pythia~is estimated to be at most $3\%$~for all \mlvjj~values.

For the remaining, smaller backgrounds modeled with MC simulation, only theoretical uncertainties due to limited knowledge of their production cross-sections are considered. The production rate of $WW$~and $ZZ$~dibosons is known 
to $5\%$~accuracy, while that for $WZ$~production is known to within $7\%$~\cite{dibosons}. The uncertainty on 
the $Z$ + jets production rate is estimated to be $5\%$, primarily due to limited knowledge of the $u$-~and $d$-quark 
PDFs~\cite{zwprod}. The production of $s$-channel single top quarks ($tb$) is known to 
$6\%$~\cite{tb} while $t$-channel ($tqb$) and $tW$~production are known to $^{+5}_{-4}\%$ 
and $9\%$~\cite{tqb_tW}, respectively.

For the signals, the PDF uncertainty is estimated by comparing signal events generated with MRST2007LO*~and CTEQ6L PDFs 
and a maximum difference of $5\%$ is measured in the acceptance. The ISR and FSR uncertainty is determined to 
be $5\%$~using the same procedure as that for \ttbar~events. 

The largest experimental uncertainties come from the determination of the jet energy scale (JES)~\cite{jesjer} and 
resolution (JER)~\cite{Aad:2012ag}. The JES uncertainty includes effects due to uncertainties in jet flavor composition, 
overlapping jets, and pile-up effects. The overall JES uncertainty on each background process as well as the signal is determined 
by varying all jet energies within their uncertainties. The impact of this uncertainty varies with \mlvjj, and the largest deviation from the nominal prediction is presented. For the background samples, this ranges from $8\%$ for single top quark events to $13\%$ for diboson events. For the signal events samples, the largest deviation from the nominal prediction for all \mlvjj~values is $4\%$. An equivalent procedure is applied to evaluate the JER uncertainty, and the largest deviation from the nominal prediction is found to be between $1\%$ and $3\%$ for all signal and background samples.

Additional uncertainties arise from the differences between data and MC simulation in the reconstruction efficiencies 
and energy or momentum resolution for electrons, muons, and \met. The electron energy scale and resolution 
uncertainties are derived by comparing $Z\rightarrow e^{+}e^{-}$~events in data and MC samples. The combined uncertainty 
is $2\mbox{--}3\%$~depending on \mlvjj. The corresponding uncertainty for muons is at most $2\%$~for any 
\mlvjj~value. The primary contribution to the \met~scale uncertainty is pile-up, but 
the impact on the \mlvjj~distribution above 500~\gev\ is less than $1\%$~for all backgrounds. The combined uncertainty on the signal acceptance ranges from $7\%$ at low \mlvjj~to $20\%$ at high \mlvjj.

The distributions from the ``fake'' lepton and $W$+jets backgrounds are normalized to the number of events in data control regions, and are therefore not affected by systematic uncertainties in the relative reconstruction efficiency in data and MC events, nor uncertainties in their respective production cross-sections. The ``fake'' lepton background normalization uncertainty is estimated by using the distributions of \met~and the scalar sum of the lepton \pt~and \met~to determine the ``fake'' lepton normalization, and quoting the maximum deviation from the \mt-fitted value. This results in an $80~(100)\%$~uncertainty on events with electrons (muons). The $W$+jets normalization uncertainty is defined as the difference between the low-\mjj~and high-\mjj~control region scale factors, resulting in an uncertainty of $9\%$.

\section{VIII. Results and Interpretation}

The numbers of expected and observed events after the final signal selection are reported in 
Table~\ref{t-signal-yields}. A total of 1453 \evjj~ and 1328 \muvjj~events are observed with background predictions of $1425\pm100$~and $1195\pm85$~events, respectively.
The \mlvjj~distributions for data, predicted background samples and an EGM~\wprime~boson signal with mass $m_{\wprime}=1$~\tev\ are shown in Fig.~\ref{f-final}. 

\begin{figure*}[b!]
  \centering
  \subfigure[]{\includegraphics[width=0.45\textwidth]{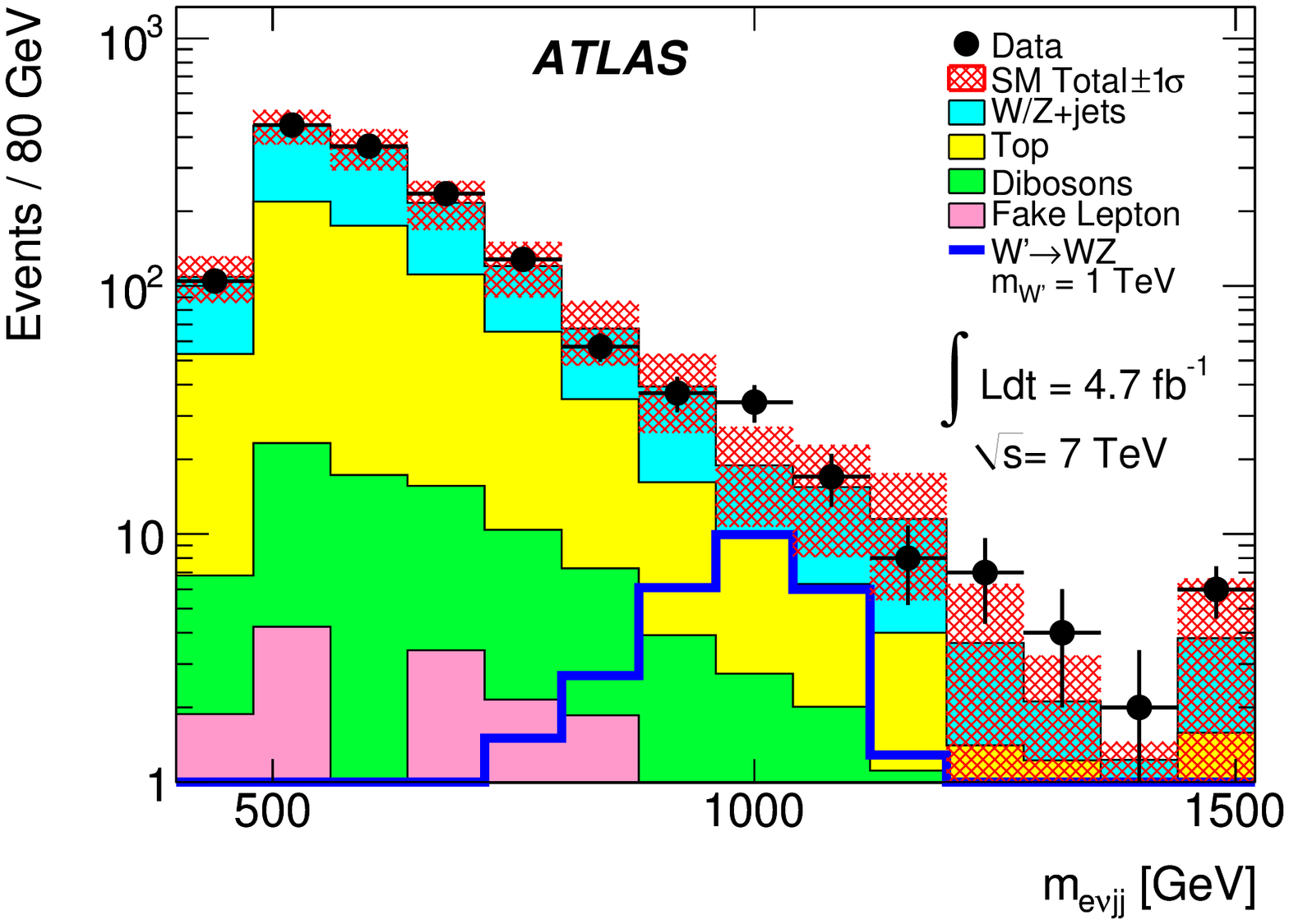}}
  \subfigure[]{\includegraphics[width=0.45\textwidth]{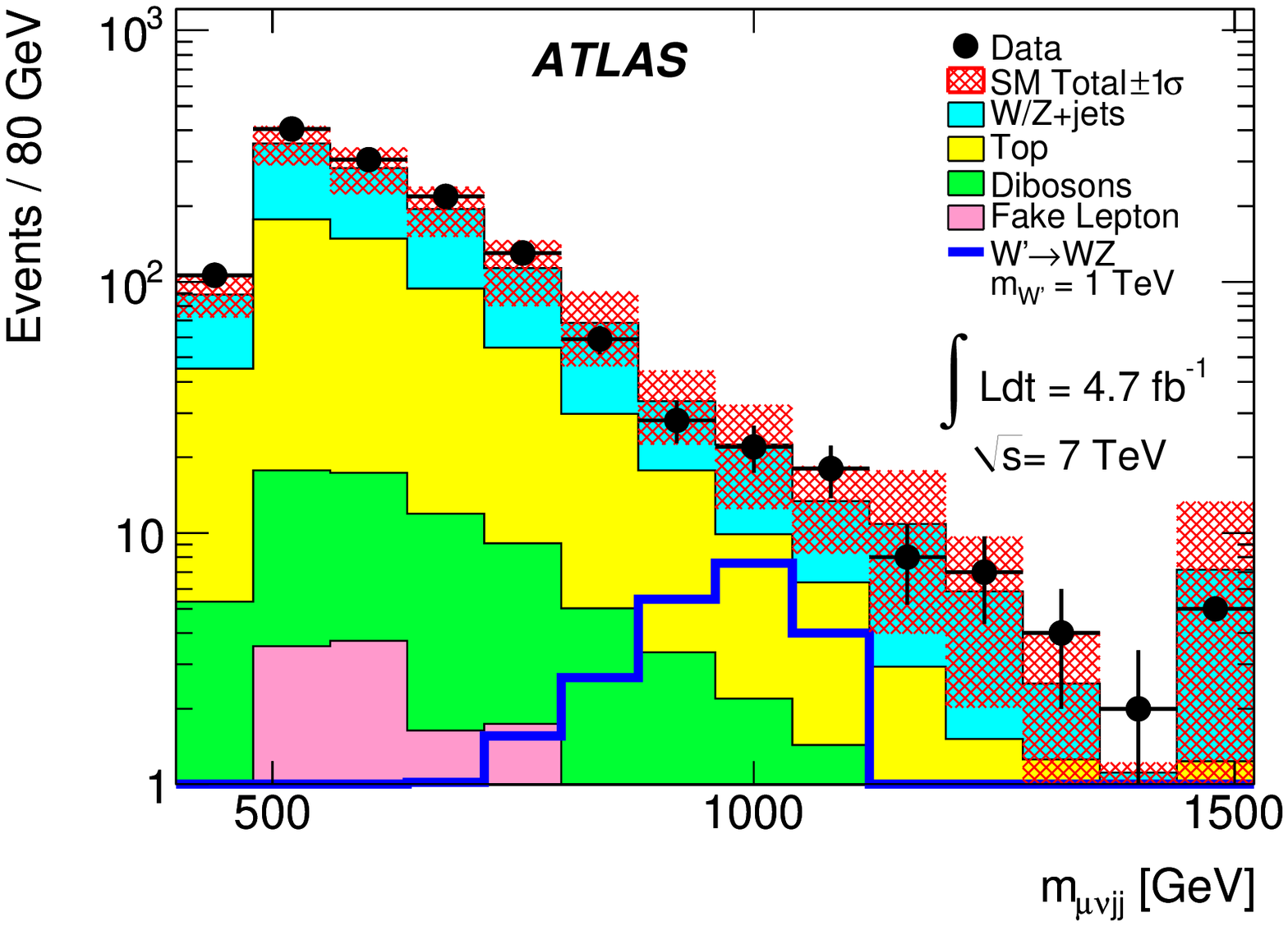}}
  \caption{(color online) Observed and predicted \mlvjj~distributions shown for all (a) \evjj~and (b) \muvjj~events satisfying the signal selection requirements. Predictions for an EGM \wprime~boson are shown for a resonance mass of 1 \tev. The right-most bin contains overflow events.
  \label{f-final}}
\end{figure*}

\begin{table}
  \caption{\label{t-signal-yields}Estimated background yields, number of data events, and predicted signal yield after applying the
signal selection criteria. Quoted uncertainties are statistical plus systematic as described in text.}
  \begin{center}
  \begin{tabular}{l | d | d}
    \hline
    \hline
      Process &\multicolumn{1}{c|}{\bf \evjj} &\multicolumn{1}{c}{\bf \muvjj} \\
    \hline
    $W$+jets & 700,65 & 590,60 \tabularnewline
    $Z$+jets & 15,2 &  15,2  \tabularnewline
    Top & 615,70 & 515,65 \tabularnewline
    Diboson & 75,9 & 60,8\tabularnewline
    Fake Lepton & 20,16 & 15,15 \tabularnewline
    Total backgrounds & 1425,100 & 1195,85 \tabularnewline
    Data & 1453 & 1328 \tabularnewline
    \hline
   RS1 $G^{*}~(m_{\graviton}=1$~\tev$)$ & 22,2 & 18,2\tabularnewline
    Bulk $G^{*}~(m_{\graviton}=1$~\tev$)$ & 4,0.4 & 3.5,0.3   \tabularnewline
    EGM $W'~(m_{\wprime}=1$~\tev$)$ & 29,2  & 24,2\tabularnewline
    \hline
    \hline

  \end{tabular}
  \end{center}
\end{table}

These distributions are used to construct a log-likelihood ratio (LLR) test statistic to compute the 
statistical significance of any excess over expectation using a modified frequentist approach.
Pseudo-experiments that treat all systematic uncertainties as Gaussian-sampled nuisance parameters are used 
to generate the distribution of possible LLR values for the background-only (b) and signal-plus-background (s+b) 
hypotheses. Confidence levels (CL) for each hypothesis are defined as the fraction of experiments with 
LLR greater than or equal to the LLR evaluated on the data.

The statistical significance of an observed signal is quantified by giving, for each mass point, the $p$-value ({\mbox{$p \equiv 1-\rm{CL}_{b}$}}) of the background-only hypothesis. The greatest deviations from the background prediction occur at \mlvjj~$=1300$~\gev~~and~$1500$~\gev~with $p=0.12$~and $0.11$, respectively. 

Lacking evidence for new phenomena, limits on the signal rate are determined using the $CL_{s}$ method~\cite{stat_junk,stat_read}. This method uses a ratio of the $p$-values of the signal-plus-background and background-only hypotheses called $CL_{s}$. For a $95\%$~CL exclusion, the signal production cross-section ($\sigma^{95\%}$) is adjusted until 
$CL_{s}=0.05$, and the resonance mass limit ($m^{95\%}$) is defined by the mass for which 
$\sigma(m^{95\%}) = \sigma^{95\%}$. The excluded production cross-sections times the branching 
ratios to the \WW~or \WZ~final state are shown in Fig.~\ref{f-limits}, with the \evjj~and \muvjj~channels combined, for the three signal hypotheses. 
The expected and observed limits on the resonances are shown in Table~\ref{t-limits}~for the
\evjj\ and \muvjj~channels separately, as well as their combination.

Limits are also set on the EGM~\wprime~boson coupling strength scaling factor $c_{\rm{EGM}}$~ within the EGM~framework. The EGM \wprime\ boson limits shown in Fig.~\ref{f-limits} correspond to $c_{\rm{EGM}}=1$. For $c_{\rm{EGM}}>10$, the resonance width exceeds the experimental resolution, thus only values less than 10 are considered. Limits on $c_{\rm{EGM}}$~are derived as a function of $m_{\wprime}$ as shown in Fig.~\ref{f-limits-2d}. 

\begin{figure*}
  \centering
  \subfigure[]{\includegraphics[width=0.45\textwidth]{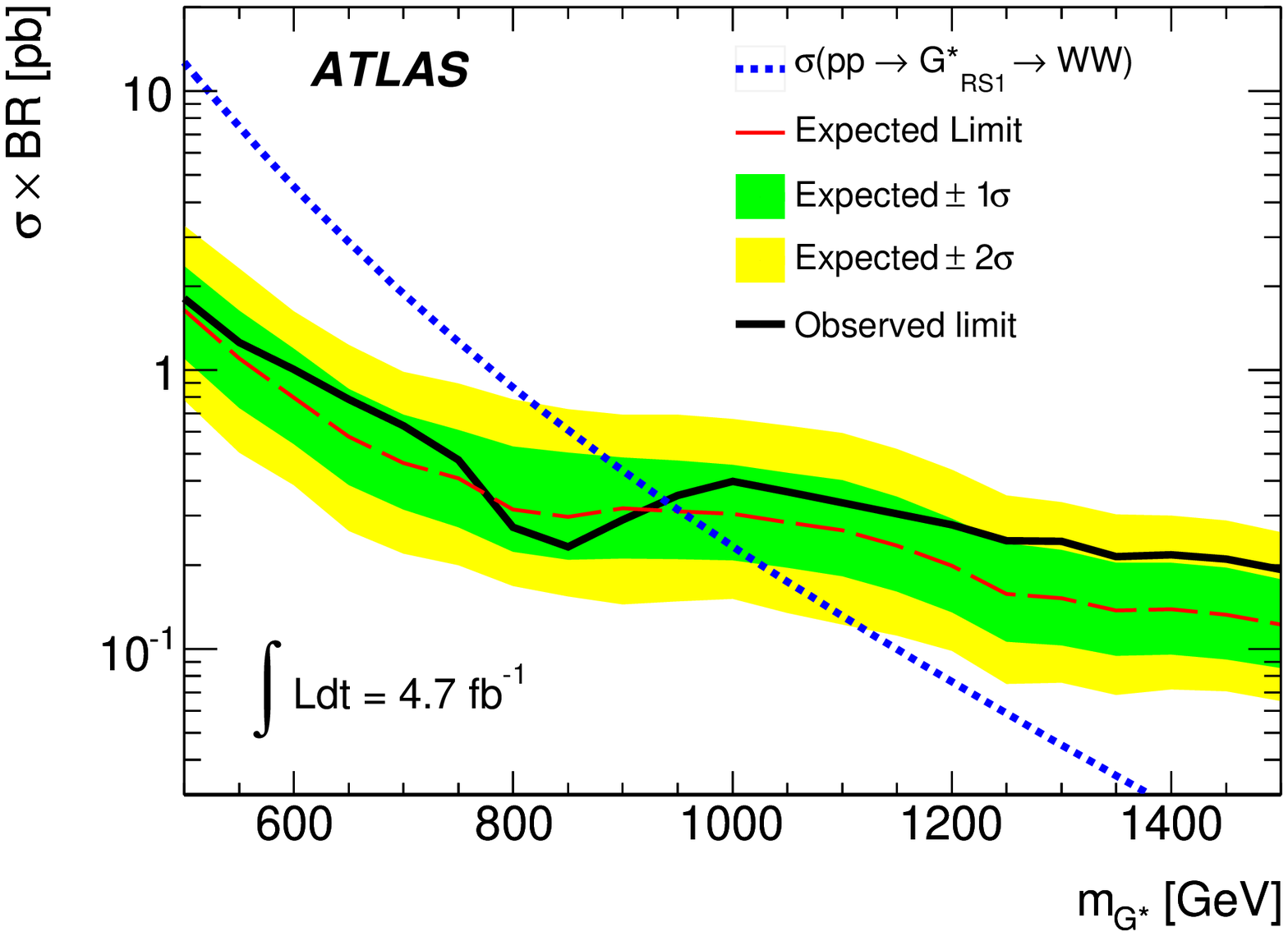}}
  \subfigure[]{\includegraphics[width=0.45\textwidth]{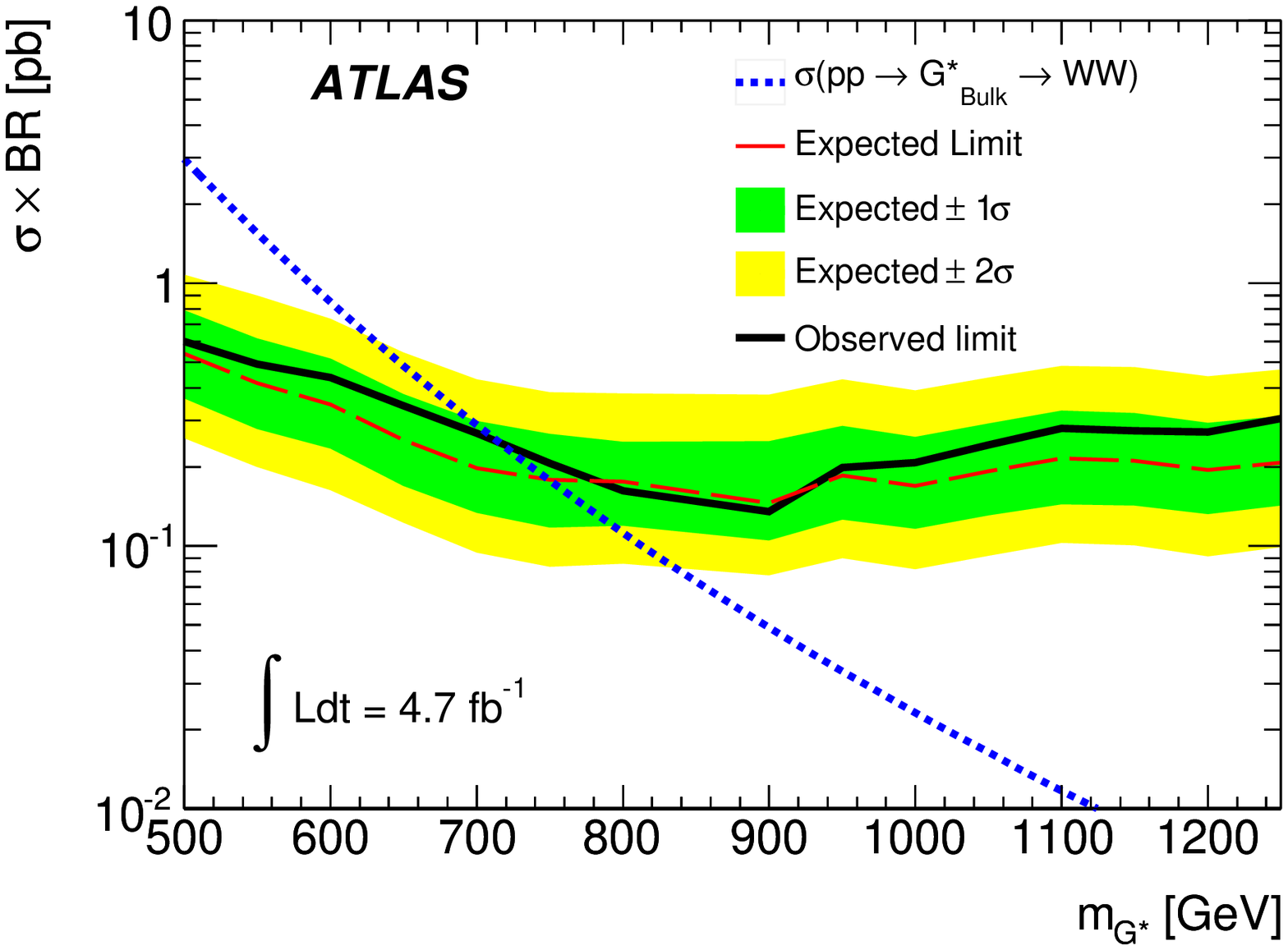}}
  \subfigure[]{\includegraphics[width=0.45\textwidth]{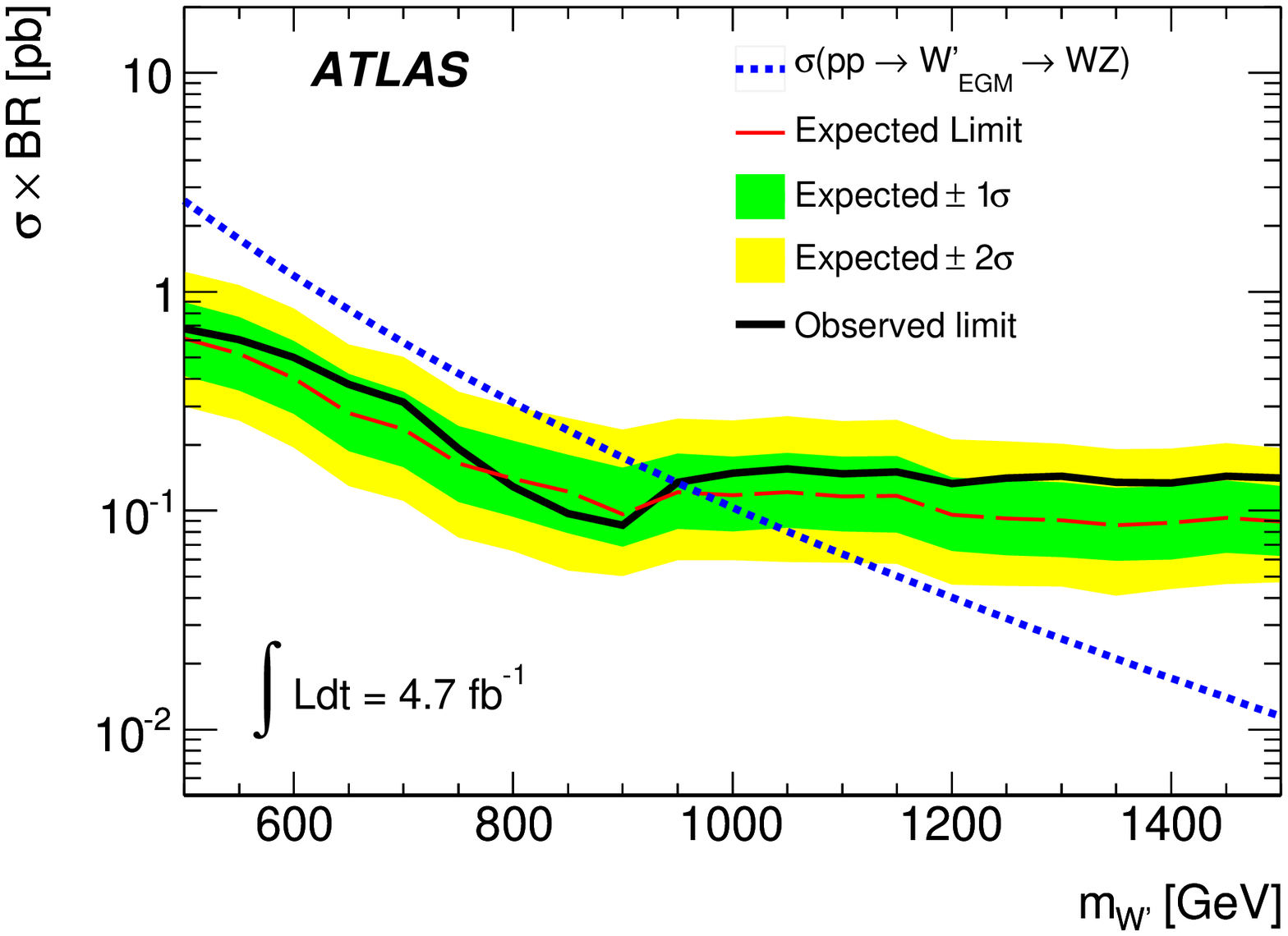}}
 \caption{
    (color online) Observed and expected $95\%$~CL upper limits on $\sigma(pp\rightarrow\graviton) \times BR(\gww)$~for (a) an RS1~\graviton and (b) a bulk RS \graviton, and on $\sigma(pp\rightarrow\wprime) \times BR(\wwz)$~for (c) an EGM \wprime~boson\label{f-limits}.}
\end{figure*}

\begin{table}
  \caption{Expected and observed $95\%$~CL lower mass limits (GeV) for the RS1 \graviton, bulk RS \graviton, and the EGM \wprime~boson using \evjj~events, \muvjj~events and the combined channels.
  \label{t-limits}}
  \begin{center}
  \begin{tabular}{l c c c}
    \hline
    \hline
    Process &\multicolumn{1}{c}{\bf \evjj} &\multicolumn{1}{c}{\bf \muvjj} &\multicolumn{1}{c}{\bf \lvjj} \\
    \hline
Expected Limits [GeV]\\
~~RS1 \graviton        & 930   &  900  &  \bf{950} \\
~~Bulk RS \graviton    & 740   &  710  &  \bf{750} \\
~~EGM \wprime          & 950   &  930  &  \bf{970} \\
\hline
Observed Limits [GeV]\\
~~RS1 \graviton        & 910   &  920  &  \bf{940} \\
~~Bulk RS \graviton    & 760   &  650  &  \bf{710} \\
~~EGM \wprime          & 930   &  930  &  \bf{950} \\
\hline
  \end{tabular}
\end{center}
\end{table}

\begin{figure*}
  \centering
    \includegraphics[width=0.5\textwidth]{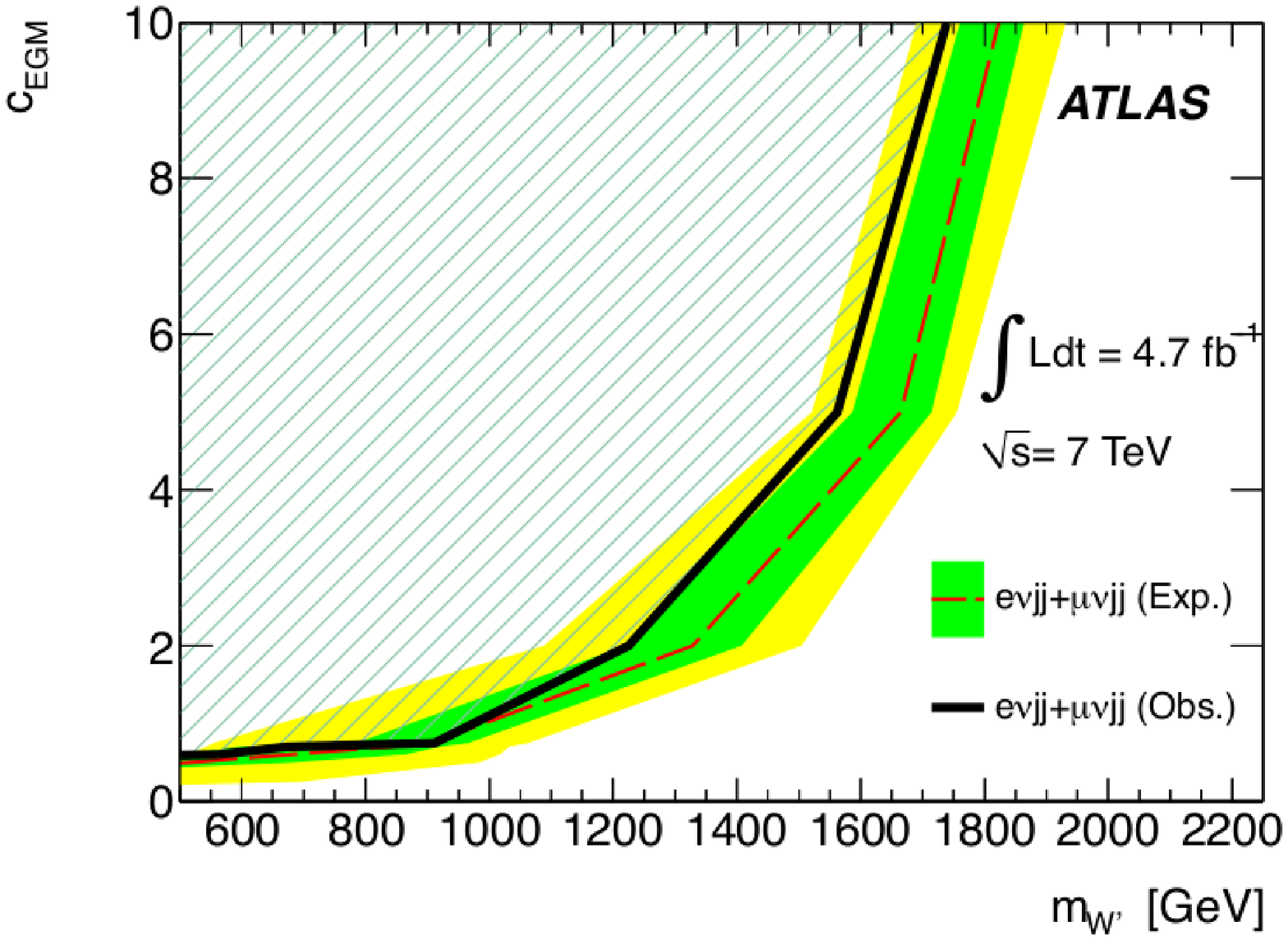}\\
  \caption{(color online) The $95$\%~CL observed and expected excluded regions of the EGM coupling strength scaling factor $c_{\rm{EGM}}$ as a function of $m_{\wprime}$. The green and yellow band correspond to the $\pm$1 and $\pm 2\sigma$ intervals, respectively.\label{f-limits-2d}}
\end{figure*}

\section{IX. Conclusion} 
\label{sec:conclusion}

We report the results of a search for resonant \WW~and \WZ~production in the \lvjj~decay channels using an integrated luminosity of \mbox{4.7 fb$^{-1}$}~of $pp$-collision data at \shat~collected in 2011 by the ATLAS detector at the Large Hadron Collider. A set of event selections for the RS1 \graviton, the bulk RS \graviton, and the EGM \wprime~boson signals are derived using simulated events. No evidence for resonant diboson production is observed and $95\%$~CL upper bounds on the two graviton and EGM~\wprime~boson production cross-sections are determined. Resonance masses below 940~\gev, 710~\gev, and 950~\gev\ are excluded at $95\%$~CL for the spin-2 RS1 graviton, the spin-2 bulk RS graviton and the spin-1 EGM \wprime~boson, respectively.

\section{Acknowledgements}

We thank CERN for the very successful operation of the LHC, as well as the
support staff from our institutions without whom ATLAS could not be
operated efficiently.

We acknowledge the support of ANPCyT, Argentina; YerPhI, Armenia; ARC,
Australia; BMWF and FWF, Austria; ANAS, Azerbaijan; SSTC, Belarus; CNPq and FAPESP,
Brazil; NSERC, NRC and CFI, Canada; CERN; CONICYT, Chile; CAS, MOST and NSFC,
China; COLCIENCIAS, Colombia; MSMT CR, MPO CR and VSC CR, Czech Republic;
DNRF, DNSRC and Lundbeck Foundation, Denmark; EPLANET, ERC and NSRF, European Union;
IN2P3-CNRS, CEA-DSM/IRFU, France; GNSF, Georgia; BMBF, DFG, HGF, MPG and AvH
Foundation, Germany; GSRT and NSRF, Greece; ISF, MINERVA, GIF, DIP and Benoziyo Center,
Israel; INFN, Italy; MEXT and JSPS, Japan; CNRST, Morocco; FOM and NWO,
Netherlands; BRF and RCN, Norway; MNiSW, Poland; GRICES and FCT, Portugal; MERYS
(MECTS), Romania; MES of Russia and ROSATOM, Russian Federation; JINR; MSTD,
Serbia; MSSR, Slovakia; ARRS and MIZ\v{S}, Slovenia; DST/NRF, South Africa;
MICINN, Spain; SRC and Wallenberg Foundation, Sweden; SER, SNSF and Cantons of
Bern and Geneva, Switzerland; NSC, Taiwan; TAEK, Turkey; STFC, the Royal
Society and Leverhulme Trust, United Kingdom; DOE and NSF, United States of
America.

The crucial computing support from all WLCG partners is acknowledged
gratefully, in particular from CERN and the ATLAS Tier-1 facilities at
TRIUMF (Canada), NDGF (Denmark, Norway, Sweden), CC-IN2P3 (France),
KIT/GridKA (Germany), INFN-CNAF (Italy), NL-T1 (Netherlands), PIC (Spain),
ASGC (Taiwan), RAL (UK) and BNL (USA) and in the Tier-2 facilities
worldwide.

\bibliographystyle{atlasBibStyleWoTitle}
\bibliography{lvjj_maintext}

\onecolumngrid
\clearpage 
\input{atlas_authlist}

\end{document}

%% file: atlas_authlist.tex
\begin{flushleft}
{\Large The ATLAS Collaboration}

\bigskip

G.~Aad$^{\rm 48}$,
T.~Abajyan$^{\rm 21}$,
B.~Abbott$^{\rm 111}$,
J.~Abdallah$^{\rm 12}$,
S.~Abdel~Khalek$^{\rm 115}$,
A.A.~Abdelalim$^{\rm 49}$,
O.~Abdinov$^{\rm 11}$,
R.~Aben$^{\rm 105}$,
B.~Abi$^{\rm 112}$,
M.~Abolins$^{\rm 88}$
% ...
\bigskip

% ...
$^{11}$ Institute of Physics, Azerbaijan Academy of Sciences, Baku, Azerbaijan\\
$^{12}$ Institut de F\'{i}sica d'Altes Energies and Departament de F\'{i}sica de la Universitat Aut\`{o}noma de Barcelona and ICREA, Barcelona, Spain\\
% ...
$^{21}$ Physikalisches Institut, University of Bonn, Bonn, Germany\\
% ...
$^{48}$ Fakult\"{a}t f\"{u}r Mathematik und Physik, Albert-Ludwigs-Universit\"{a}t, Freiburg, Germany\\
$^{49}$ Section de Physique, Universit\'{e} de Gen\`{e}ve, Geneva, Switzerland\\
% ...
$^{88}$ Department of Physics and Astronomy, Michigan State University, East Lansing MI, United States of America\\
% ...
$^{105}$ Nikhef National Institute for Subatomic Physics and University of Amsterdam, Amsterdam, Netherlands\\
% ...
$^{111}$ Homer L. Dodge Department of Physics and Astronomy, University of Oklahoma, Norman OK, United States of America\\
$^{112}$ Department of Physics, Oklahoma State University, Stillwater OK, United States of America\\
% ...
$^{115}$ LAL, Universit\'{e} Paris-Sud and CNRS/IN2P3, Orsay, France\\
% ...
\end{flushleft}


%% file: lvjj_maintext.bbl
\begin{thebibliography}{10}
\bibitem{geometry}
ATLAS uses a right-handed coordinate system with its origin at the nominal
  interaction point~(IP) in the center of the detector and the $z$-axis along
  the beam pipe. The $x$-axis points from the IP to the center of the LHC ring,
  and the $y$-axis points upward. Cylindrical coordinates $(r,\phi)$ are used
  in the transverse plane, $\phi$ being the azimuthal angle around the beam
  pipe. The pseudorapidity is defined in terms of the polar angle $\theta$ as
  $\eta=-\ln\tan(\theta/2)$. The transverse energy $E_{\rm T}$ is defined as $E
  \sin{\theta}$, where $E$ is the energy associated to the calorimeter cell or
  energy cluster. Similarly, \pt\ is the momentum component transverse to the
  beam line.
\end{thebibliography}
